\documentclass[aps,prb,twocolumn,superscriptaddress,showpacs,amsmath,floatfix]{revtex4}
\usepackage{graphicx}
\usepackage{dcolumn}
\usepackage{multirow}

\usepackage{amsmath}
\usepackage{subfigure}
\usepackage{braket}
\usepackage{color}
\usepackage{epstopdf}
\usepackage[normalem]{ulem}

\begin{document}
\title{Ab initio theory and modeling of water}

\author{Mohan Chen}
  \affiliation{Department of Physics, Temple University, Philadelphia, Pennsylvania 19122, USA}
\author{Hsin-Yu Ko}
  \affiliation{Department of Chemistry, Princeton University, Princeton, New Jersey 08544, USA}
\author{Richard C. Remsing}
  \affiliation{Department of Chemistry, Temple University, Philadelphia, Pennsylvania 19122, USA}
  \affiliation{Institute for Computational Molecular Science, Temple University, Philadelphia, Pennsylvania 19122, USA}
\author{Marcos F. Calegari Andrade}
  \affiliation{Department of Chemistry, Princeton University, Princeton, New Jersey 08544, USA}
\author{Biswajit Santra}
  \affiliation{Department of Chemistry, Princeton University, Princeton, New Jersey 08544, USA}
\author{Zhaoru Sun}
  \affiliation{Department of Physics, Temple University, Philadelphia, Pennsylvania 19122, USA}
\author{Annabella Selloni}
  \affiliation{Department of Chemistry, Princeton University, Princeton, New Jersey 08544, USA}
\author{Roberto Car}
  \affiliation{Department of Chemistry, Princeton University, Princeton, New Jersey 08544, USA}
\author{Michael L. Klein}
  \email{mlklein@temple.edu}
  \affiliation{Department of Physics, Temple University, Philadelphia, Pennsylvania 19122, USA}
  \affiliation{Department of Chemistry, Temple University, Philadelphia, Pennsylvania 19122, USA}
  \affiliation{Institute for Computational Molecular Science, Temple University, Philadelphia, Pennsylvania 19122, USA}
\author{John P. Perdew}
  \affiliation{Department of Physics, Temple University, Philadelphia, Pennsylvania 19122, USA}
  \affiliation{Department of Chemistry, Temple University, Philadelphia, Pennsylvania 19122, USA}
\author{Xifan Wu}
  \email{xifanwu@temple.edu}
  \affiliation{Department of Physics, Temple University, Philadelphia, Pennsylvania 19122, USA}
  \affiliation{Institute for Computational Molecular Science, Temple University, Philadelphia, Pennsylvania 19122, USA}
\date{\today }

\begin{abstract}
Water is of the utmost importance for life and technology.
However, a genuinely predictive {\it ab initio} model of water has eluded scientists.
We demonstrate that a fully {\it ab initio} approach,
relying on the strongly constrained and appropriately normed (SCAN) density functional,
provides such a description of water.
SCAN accurately describes the balance among covalent bonds, hydrogen bonds, and van der Waals
interactions that dictates the structure and dynamics of liquid water.
Notably, SCAN captures the density difference between water and ice I{\it h} at ambient conditions,
as well as many important structural, electronic, and dynamic properties of liquid water.
These successful predictions of the versatile SCAN functional
open the gates to study complex processes in aqueous phase chemistry and the interactions of water
with other materials in an efficient, accurate, and predictive, {\it ab initio} manner.

Keywords: water, ab initio theory, hydrogen bonding, density functional theory, molecular dynamics.
\end{abstract}

\maketitle

\section{Significance}
Water is vital to our everyday life, but its structure at a molecular level is still not fully understood
from {either experiment or theory.
The latter is hampered} by our inability to construct a purely predictive, first principles model.
The difficulty in modeling water lies in capturing the delicate interplay among the many strong and weak forces that govern its behavior and phase diagram.
Herein, molecular simulations with a recently proposed non-empirical quantum mechanical approach (the SCAN density functional), yield an excellent description of the structural, electronic, and dynamic properties of liquid water.
SCAN-based approaches, which describe diverse types of bonds in materials on an equal, accurate footing, will likely enable efficient and reliable modeling of aqueous phase chemistry.

\newpage
Water is arguably the most important molecule for life
and is involved in almost all biological processes.
Without water,
life, as we know it, would not exist,
earning water the pseudonym {\it matrix of life}, among others~\cite{Ball:2008}.
Despite the apparent simplicity of an H$_2$O molecule, water in the condensed phase
displays a variety of anomalous properties which originates from its complex
structure.
%
In an ideal arrangement, water molecules form a tetrahedral network of hydrogen (H) bonds
with each vertex being occupied by a water molecule.
This tetrahedral network is realized in the solid phase ice I{\it h}, but thermal fluctuations
disrupt the H-bond network in the liquid state, with the network fluctuating on picosecond to nanosecond timescales.
Due to the complexity of the H-bond network and its competition with thermal fluctuations,
a precise molecular-level understanding of the structure of liquid water remains elusive.
Major challenges lie in unambiguously capturing the atomic-scale fluctuations
in water experimentally.
Current approaches such as time-resolved spectroscopy~\cite{Fecko:Science:2003,04S-Wernet}
and diffraction measurements~\cite{08L-Soper,13JCP-Skinner} may be able to resolve changes on picosecond
timescales, but rely on interpretation through models, which often cannot describe all the details of liquid
water with quantitative accuracy.
Not surprisingly, the nature of the H-bond network in liquid water continues to
be at the center of scientific debate
and advances in both experiment and theory are needed, especially with regard
to quantitative modeling of aqueous phase chemistry.

{\it Ab initio} molecular dynamics (AIMD) simulation~\cite{85L-CPMD} is an ideal approach
for modeling the condensed phases of water
across the phase diagram and aqueous phase chemistry
using quantum mechanical principles~\cite{09JCTC-Kuhne,13JCP-Biswajit,14JCP-Rob,16-Gillan,15JPCL-Alex},
although for some applications, such as the study of liquid vapor phase equilibria~\cite{06JPCA-McGrath}, Monte Carlo methods are better suited.
In particular, Kohn-Sham density functional theory (DFT)~\cite{kohn65} --- used to model the system in its electronic ground state ---
provides an efficient framework that enables the simulation of the length and time scales needed to converge
many statistical mechanical averages in disordered, liquid state systems.
The DFT formalism is exact for the electronic ground-state energy and density,
but in practice approximations must be adopted to describe many-body effects,
included in the exchange-correlation (XC) functional.
%
XC functionals can be conceptually arranged, by accuracy and computational efficiency,
according to Jacob's ladder~\cite{perdew2001jacob},
with the simplest local density approximation (LDA)~\cite{80L-Ceperley,81B-Perdew} on the bottom rung of the ladder,
followed by generalized gradient approximations (GGAs)~\cite{88A-Becke,88B-Lee,96L-PBE},
meta-GGAs, hybrid functionals~\cite{96JCP-Perdew,99JCP-Adamo}, and so on.
The past three decades have witnessed widespread successes of DFT
in elucidating and predicting properties of materials.
However, water still presents a major challenge, with many DFT-based simulations yielding
results that are not even qualitatively consistent with experimental measurements.
%
The H-bonds formed between gas-phase water clusters were first treated within the LDA~\cite{la92,la93},
which overestimates H-bond strengths and yields inter-water distances that are too close.
This overbinding is largely corrected by GGA-level functionals,
which became a class of popular functionals to study liquid water within the last two decades
~\cite{16-Gillan}.
Despite the improvements over LDA that are provided by GGAs,
H-bond strengths are overestimated and, consequently,
the dynamical properties predicted by GGAs are generally much too slow.
%
Worse still, GGAs predict that ice sinks in water, that is, water has a lower density than ice~\cite{15JPCL-Alex,09JPCB-Schmidt,11JCP-Wang,15JCP-Miceli,15JPCL-Alex}.
These disagreements remain even after considering hybrid functionals~\cite{15JPCL-Alex}
and accounting for nuclear quantum effects (NQEs)~\cite{13PNAS-Ceriotti}, illustrating that the deficiencies
are a manifestation of errors within the underlying GGA to the XC functional.

The difficulty in modeling liquid water with DFT arises from the delicate nature of the H-bond network.
A H-bond is a directional attractive force between
the oxygen of one molecule and the protons of another.
While mainly electrostatic in nature,
H-bonds also exhibit a non-negligible covalency.
Notably, a covalent O-H bond binds one order of magnitude stronger than a H-bond in water.
Therefore, a slightly misbalanced covalent bond inevitably incurs a
non-negligible error in the predicted H-bond strength.
Moreover, water molecules interact with each other through van der Waals (vdW) dispersion forces at larger distances,
which are non-directional and in general weaker than H-bonds by roughly an order of magnitude.
Thus, one needs to capture the balance among interactions whose magnitudes vary
by orders of magnitude in water.
The short-ranged portion of the vdW interactions have been captured by local and semi-local XC functionals.
In contrast, the intermediate- and long-ranged parts of the vdW interactions have not been captured
by any general-purpose GGA.
Recent studies have identified vdW interactions as an important determinant of water structure;
vdW interactions often lead to more disordered water structures,
more accurate water densities, and improved dynamic properties
~\cite{09JPCB-Schmidt,11JCP-Wang,Baer:2011rz,11JSP-Remsing,15JCP-Miceli,15JPCL-Alex,15JCP-Ben}.
Thus, the H-bond network of liquid water is produced by a delicate competition among covalent bonds,
H-bonds, and vdW interactions, and describing this complex interplay of interactions
continues to be a highly challenging task.
In this regard, non-empirical, general purpose XC functionals that describe all types of interactions
on an equal footing are imperative but still largely absent in the literature.

To address the above issues,
we performed AIMD simulations of liquid water in the isothermal-isobaric ensemble~\cite{80L-Parrinello},
employing the strongly constrained and appropriately normed (SCAN)
meta-GGA functional~\cite{15L-Sun}.
SCAN is inherently non-empirical,
developed by satisfying all 17 known exact constraints on semi-local XC functionals.
Thus, the results obtained from SCAN are purely predictive and do not rely on training data.
SCAN was shown to predict the energetics of gas-phase water hexamers
and ice phases with quantitative accuracy, while other XC functionals,
even with vdW corrections, were unable to make even qualitative predictions~\cite{16NC-Sun}.
This suggests that SCAN possesses the ingredients necessary to describe liquid water.
Indeed, we demonstrate that SCAN predicts
structural, electronic, and dynamic properties of liquid water in excellent agreement with experimental measurements.
In particular, due to its ability to describe vdW interactions on intermediate length-scales,
SCAN yields the correct density ordering between liquid water and ice,
correctly predicting that ice floats on liquid water.
The dynamics of liquid water are also improved to near quantitative agreement with experiments.
We expect the computationally-efficient and accurate SCAN functional to serve as a major quantum mechanics-based tool
for studying chemical processes in aqueous media.

\section{Molecular and Electronic Structure of Liquid Water}
The pair structure of liquid water can be measured by X-ray diffraction~\cite{08L-Soper,13JCP-Skinner}
and neutron diffraction experiments~\cite{08L-Soper},
from which structural information is contained in the resulting radial distribution functions (RDFs).
We compare the RDFs
obtained from AIMD simulations with SCAN and the Perdew-Burke-Ernzerhof (PBE)~\cite{96L-PBE} GGA,
as well as the experimental data.
{Here we compare two fully {ab initio} density functionals, without an empirical dispersion (D) correction to either.
While such a correction improves PBE for solids and liquids~\cite{14JPCC-Arindam}, it slightly worsens PBE's unacceptable overbinding of molecules, and thus PBE-D is not recommended for reactions in solvents.}
Figs.~1(a) and~1(b) show the oxygen-oxygen and oxygen-hydrogen RDFs,
g$_{\mathrm{OO}}$(r) and g$_{\mathrm{OH}}$(r), respectively.
SCAN dramatically improves almost all
features in g$_{\mathrm{OO}}$(r) and g$_{\mathrm{OH}}$(r),
producing a pair structure in much better agreement with experimental measurements than PBE.
The first peak of g$_{\mathrm{OH}}$(r) contains all correlations within the covalent O-H bonds.
SCAN enhances the covalency of water molecules,
shortening the covalent bond length to 0.977~\AA~(first maximum in g$_{\mathrm{OH}}$(r)),
in comparison to the 0.989~\AA~from PBE.
The shorter O-H bond length indicates that the
oxygen and protons bind more strongly.
Consequently, the protons of water molecules are less easily donated to form H-bonds.

\begin{figure}[tb]
\label{fig1}
\begin{center}
\includegraphics[width=0.35\textwidth]{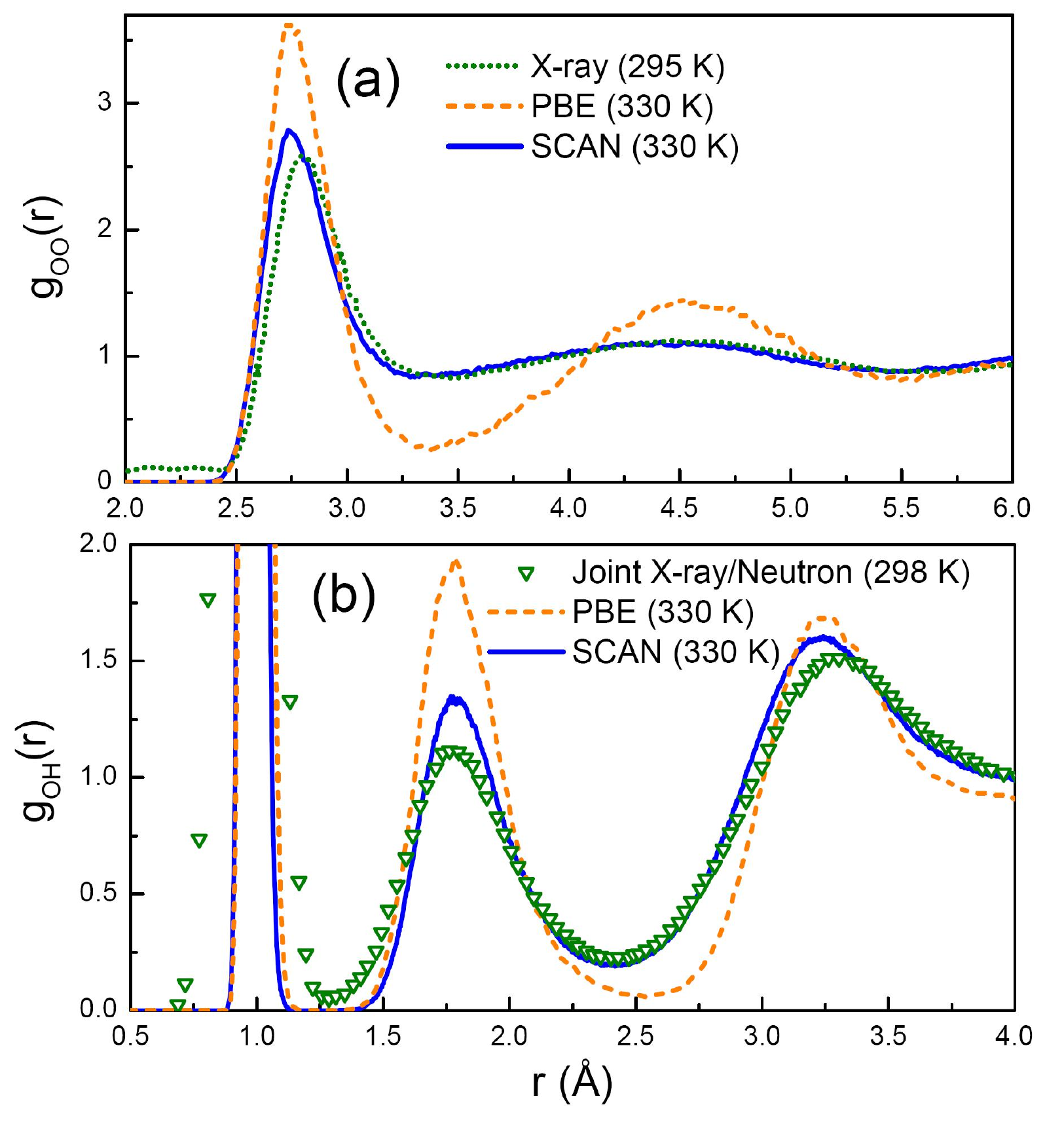}
\end{center}
\caption
{Radial distribution functions (a) g$_{\mathrm{OO}}$(r) and (b) g$_{\mathrm{OH}}$(r)
of liquid water predicted by PBE and SCAN at 330 K, as well as
that from X-ray diffraction experiments~\cite{13JCP-Skinner} for g$_{\mathrm{OO}}$(r)
and joint X-ray/neutron diffraction experiments~\cite{08L-Soper} for g$_{\mathrm{OH}}$(r). An elevated
temperature of 30 K was utilized in AIMD simulations to mimic NQEs~\cite{08L-Morrone}.
}
\end{figure}

%
Correlations between H-bonded neighbors are contained in the first peak of g$_{\mathrm{OO}}$(r)
and the second peak of g$_{\mathrm{OH}}$(r).
As evidenced by Fig.~1, SCAN captures these correlations with high accuracy
due to its ability to describe H-bonding.
The region between the first and second peaks of g$_{\mathrm{OO}}$(r) predominantly consists of
non-H-bonded water molecules that occupy the interstitial space between H-bonded neighbors;
the increased number of water molecules in the interstitial regions is due to vdW interactions,
as discussed further below.
Subsequent coordination shells are also captured by SCAN, evidenced by the good agreement between the second and third peaks in g$_{\mathrm{OO}}$(r).
We emphasize that the near perfect agreement between the SCAN g$_{\mathrm{OO}}$(r) and experiment
is non-trivial, because the structure of water is a manifestation of
the delicate interplay among covalent bonds, H-bonds, and vdW interactions.
%

\begin{table*}[tbh]
\caption{
Properties of water (330 K) and ice I{\it h} (273 K) predicted by SCAN and PBE functionals
in the isobaric-isothermal ensemble:
densities of water ($\rho_{w}$)
and ice I{\it h} ($\rho_{{\rm I}h}$),
density difference ($\Delta \rho$),
density ratio $\rho_{w}/\rho_{{\rm I}h}$,
{dipole moments of water ($\mu_{w}$)
and ice I{\it h} ($\mu_{{\rm I}h}$)},
band gap ($E_{g}$),
tetrahedral order parameter ($q$),
diffusion coefficient ($D$),
and rotational correlation time ($\tau_2$).
{The temperatures for experimental data (EXP) $\rho_{w}$,
$\rho_{{\rm I}h}$, $\mu_{w}$, $D$, and $\tau_2$
are 300~\cite{NIST}, 273~\cite{CRC}, 298~\cite{00JCP-Badyal}, 298~\cite{73JPC-Mills}, and 300 K~\cite{01JACS-Ropp}, respectively.
The experimental $q$ value~\cite{08L-Soper}
was obtained by combining X-ray diffraction at 296 K and
neutron diffraction data at 298 K in a structural model using empirical potential structural refinement.
No experimental data of $\mu_{Ih}$ are found but an induction model
gave rise to 3.09 D for $\mu_{Ih}$~\cite{98JCP-Batista}.}
Experimental data for $q$, $D$, and $\tau_2$ are for D$_2$O
chosen for consistency with the masses used in simulations for the dynamic properties.
Error bars correspond to one standard deviation.}
\scalebox{1.00}{
\begin{tabular}{cccccccccccc}
  \hline
  \hline
  Method &  $\rho_{w}$ (g/mL) & $\rho_{{\rm I}h}$ (g/mL) & $\Delta\rho$ (g/mL) & $\rho_{w}/\rho_{{\rm I}h}$ &
  $\mu_{w}$ (D) & {$\mu_{Ih}$ (D)} & $E_{g}$ (eV) & $q$ & $D$ (\AA$^2$/ps) & $\tau_2$ (ps) \\
  \hline
  SCAN & 1.050$\pm$0.027 & 0.964$\pm$0.023 &  0.086{$\pm$0.035}  & 1.089{$\pm$0.038} & 2.97$\pm$0.29 & {3.29$\pm$0.21} & 4.92$\pm$0.14 & 0.68$\pm$0.18 & {0.190$\pm$0.025} & 2.9$\pm$0.4 \\
  PBE  & 0.850$\pm$0.016 & 0.936$\pm$0.013 & -0.086{$\pm$0.021} & 0.908{$\pm$0.021} & 3.12$\pm$0.28 & {3.35$\pm$0.21} & 4.43$\pm$0.13 & 0.83$\pm$0.11 & {0.018$\pm$0.002} & 7.1$\pm$0.5 \\
  EXP  & {0.99656}~\cite{NIST}  &  {0.9167}~\cite{CRC}  & {0.080}  & {1.087} & 2.9$\pm$0.6~\cite{00JCP-Badyal} &
  & 8.7$\pm$0.6~\cite{97CP-Bernas} & {0.593}~\cite{08L-Soper} & {0.187}~\cite{73JPC-Mills}  & 2.4~\cite{01JACS-Ropp} \\
  \hline
  \hline
\end{tabular}
}
\end{table*}

%
The strength of directional H-bonds is largely determined by the electronic structure of water molecules.
%
The electronic density of states (DOS) of liquid water, averaged over trajectories, is shown in Fig.~2(a)
and compared to the DOS measured by full valence
band photoemission spectroscopy~\cite{04JPCA-Winter}.
The four peaks of the DOS are assigned to the
$2a_1$, $1b_2$, $3a_1$, and $1b_1$ orbitals based on the spatial symmetries of the water molecule.
The simulated DOS are aligned at the position of the $1b_1$ orbital peak~\cite{16L-WeiChen} .
The energy difference between the $2a_1$ peak predicted by PBE and experiment is 2.3 eV.
SCAN substantially lowers this energy difference to 0.9 eV, providing a much better
description of the strongly bound $2a_1$ orbital than the GGA-level description provided by PBE.
Note that the strongly bound  $2a_1$ orbital is mainly composed of
the characteristic $2s$ orbital and is close to the oxygen atom.

\begin{figure}[tb]
\label{fig2}
\begin{center}
\includegraphics[width=0.35\textwidth]{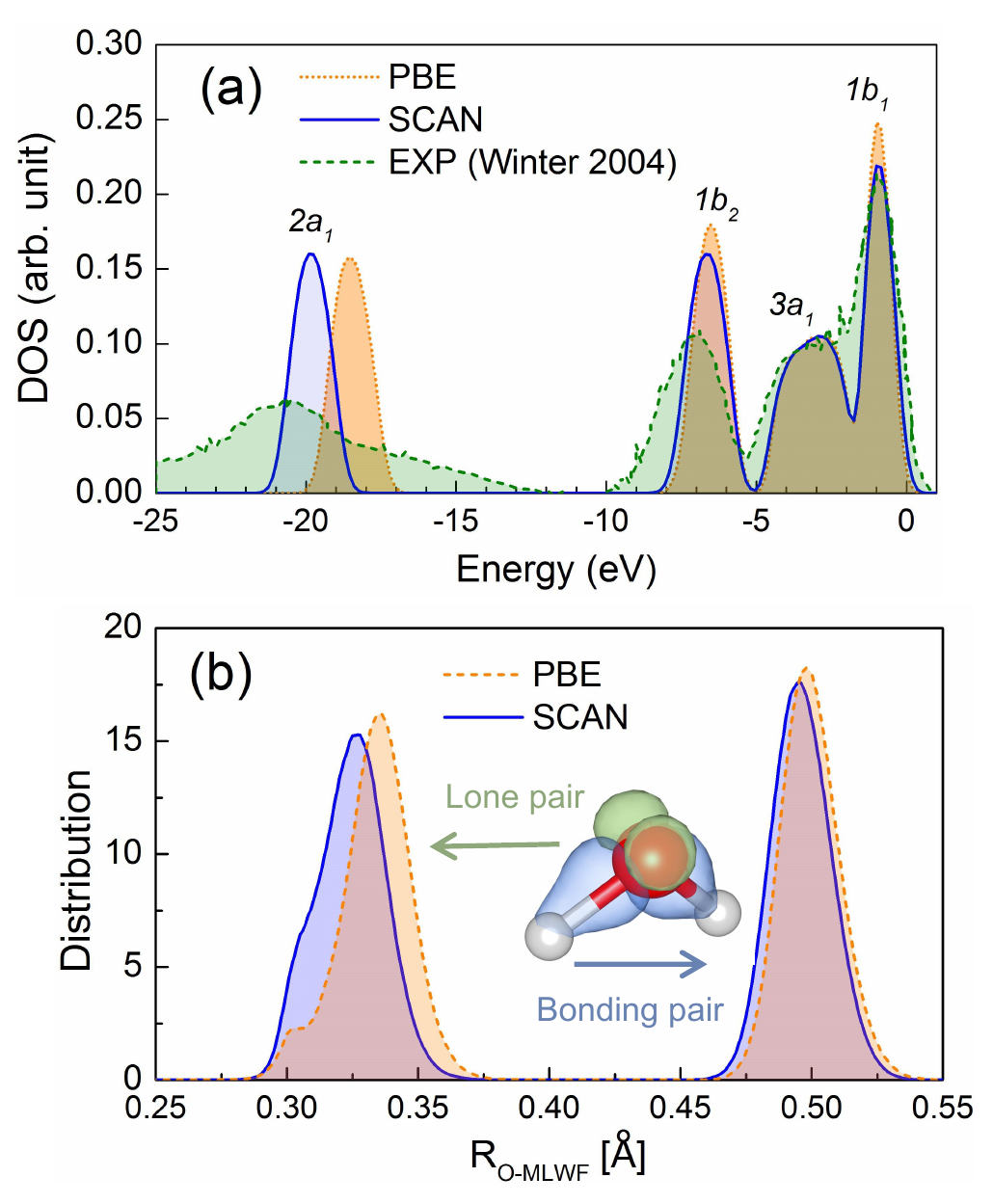}
\end{center}
\caption
{(a) Density of states (DOS) of liquid water, averaged over SCAN and PBE trajectories,
as well as from photoemission spectroscopy~\cite{04JPCA-Winter}.
The peaks are labeled according to the symmetric orbitals of a water molecule with $C_{2v}$ symmetry.
Data are aligned~\cite{16L-WeiChen} to the $1b_1$ peak of the experimental (EXP) data.
(b) Distributions of the centers of maximally localized Wannier functions (MLWFs) with respect to the oxygen position
for lone and bonding electron pairs.
The inset shows a representative snapshot of the MLWFs of a water molecule;
lone and bonding pair MLWFs are colored green and blue, respectively.
}
\end{figure}

%
The above four orbitals are related
to the two lone electron pairs and
two bonding electron pairs of a water molecule;
the lone electron pairs are closely connected to
the $2a_1$ and $1b_1$ orbitals while the bonding electron pairs
have a strong relation to the $1b_2$ and $3a_1$ orbitals.
Therefore, the improved DOS by SCAN implies that the lone and bonding electron
pairs are better captured than those from PBE.
We examine the lone and bonding electron pairs on an equal footing
through maximally localized Wannier functions (MLWFs)~\cite{12MLWF},
which are generated from a unitary transformation of the occupied Kohn-Sham eigenstates.
Fig.~2(b) shows the distributions of the centers of the MLWFs.
The lone electron pairs are closer to the oxygen atom in the SCAN description of water
than PBE, while the bonding electron pairs only differ slightly between the two XC functionals.
The smaller distance between lone electron pairs and oxygen in SCAN
leads to a less negative environment around the lone electron pairs and
explains the lower energy of the $2a_1$ orbital in comparison to that of PBE.
Meanwhile, the nearly unchanged description of bonded electron pairs in the two functionals
is consistent with the observation that $1b_2$ and $3a_1$ states are also similar.
Consequently, electrostatic attractions between oxygen nuclei and
protons of neighboring water molecules are weaker in SCAN than PBE,
weakening the directional H-bond strength.
In addition to improving the intermolecular structure, the reduced H-bond strength
in SCAN also improves the intramolecular structure of water.
The shorter distance between the lone electron pairs and the oxygen nucleus
weakens the capability to accept H-bonds and water molecules become less polarizable.
The reduction in polarizability is expected to improve
other electronic properties of liquid water, moving them in closer agreement with experimental measurements.
Indeed, the dipole moment $\mu$ of liquid water,
computed via MLWFs, is reduced by SCAN.
Table~1 shows that $\mu=3.12$~D with PBE, while $\mu$ reduces to 2.97~D with SCAN,
in {better agreement} with experimental measurements of 2.9$\pm$0.6~D~\cite{00JCP-Badyal}.
This improvement indicates that the important dipole-dipole interactions in liquid water
are better described by SCAN.
We also estimate the band gap of water, $E_{g}$, by averaging over eight randomly selected configurations
from the trajectories.
SCAN and PBE predict $E_g=4.92$ and 4.43~eV, respectively.
While SCAN improves $E_g$ by about 0.5~eV,
it differs significantly from the experimental value of 8.7~eV~\cite{97CP-Bernas}.
We attribute this discrepancy to the well-known underestimation
of band gaps by GGAs and meta-GGAs.

\begin{figure*}[tb]
\label{fig4}
\begin{center}
\includegraphics[width=0.75\textwidth]{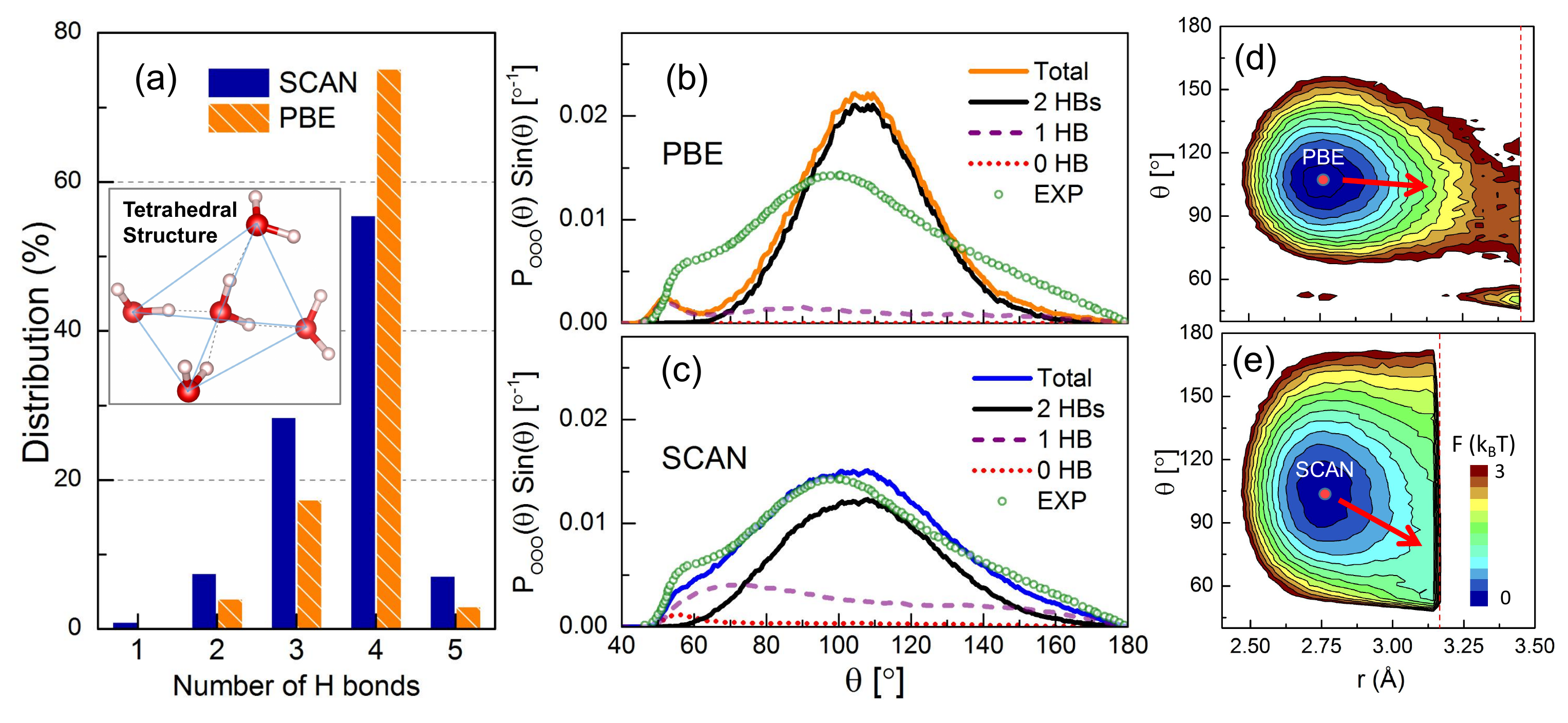}
\end{center}
\caption
{(a) Distributions of the number of hydrogen bonds in liquid water from SCAN and PBE. The inset
illustrates an ideal tetrahedral H-bonding structure.
Oxygen and hydrogen atoms are respectively depicted in red and white; H-bonds
are shown with dashed lines.
(b,c) Bond angle distributions $P_{\mathrm{OOO}}(\theta)$ from (b) PBE and (c) SCAN.
$P_{\mathrm{OOO}}(\theta)$
is decomposed into contributions arising from waters with a fixed number of HBs (2, 1, and 0) between
a central oxygen and its two nearest neighbors.
The experimental $P_{\mathrm{OOO}}$ of D$_2$O is inferred from experiments~\cite{08L-Soper},
and the area of $P_{\mathrm{OOO}}$ is normalized to unity.
(d,e) Free energies ($F$) as a function of $\theta$ and the oxygen-oxygen distance $r$ from (d) PBE and (e) SCAN.
The free energy minimum is identified by the red circle and referenced to zero.
The direction of change of the free energy minimum with increasing $r$ is shown with a red arrow.
The cutoff distance
used for computing the free energies is the same as that for $P_{\mathrm{OOO}}$ and is shown with a dashed red line.
}
\end{figure*}

%
The SCAN functional can describe the intermediate-ranged vdW interactions~\cite{16NC-Sun},
which shift the first minimum and the second maximum of $g_{\mathrm{OO}}$(r)
toward the first peak, with respect to that of PBE (without vdW interactions), as shown in Fig.~1(a).
Water molecules beyond the first coordination shell experience
non-directional attractions from surrounding water molecules in SCAN
and
are pulled into the interstitial spaces between H-bonded waters by these vdW forces.
Consequently, the peak position of the second coordination
shell shifts inward towards the central oxygen,
and the population of interstitial waters increases,
illustrated by the increase in the height of the first minimum in $g_{\mathrm{OO}}$(r).
Thus, the inclusion of non-directional vdW interactions on intermediate length-scales
leads to a more disordered and highly-packed water structure.
From the increased packing,
one expects the density of liquid water predicted by SCAN to be larger than that from PBE.
Moreover, the dominant effect of vdW interactions
is to provide cohesive interactions between molecules in condensed phases.
Within the vdW picture of liquids,
this leads to a cohesive pressure of magnitude $-a\rho_w^2$,
which ``squeezes'' water molecules closer together~\cite{11JSP-Remsing};
$a$ is the vdW constant and a measure of the strength of these attractive interactions,
and $\rho_w$ is the density of liquid water.
Indeed, the SCAN functional predicts $\rho_{w}$ to be
significantly higher than that predicted by PBE, as shown in Table~1.
Another problem of paramount importance is that solid water,
ice I{\it h}, floats on liquid water near ambient conditions.
This is probably the most widely known anomalous property of water.
Yet, almost all DFT-based approaches,
except some of those relying on empirical parameters,
predict a solid phase that is denser than the liquid.
In this regard, we also carried out AIMD simulations of ice I{\it h} containing 96 water molecules
at 273~K.
SCAN predicts a $\rho_{w}$ that is
larger than the density of ice I{\it h} ($\rho_{{\rm I}h}$), Table~1,
while for PBE, $\rho_{{\rm I}h}>\rho_{w}$.
The water density from SCAN is $\simeq5\%$ larger than that determined experimentally,
which is a significant improvement over the 15\%, 25\%, and 39\% underestimation
by the PBE, BLYP~\cite{88A-Becke,88B-Lee,09JPCB-Schmidt}, and PBE0~\cite{15JPCL-Alex}
functionals, respectively.
Compared to PBE,
SCAN increases $\rho_{w}$ and $\rho_{{\rm I}h}$ by 21\% and 3\%, respectively.
The 21\% increase of $\rho_w$ by SCAN
is vital in correcting the density ordering between the two phases by other functionals.
Indeed, the experimental density difference 
between liquid water and ice I{\it h},
$\Delta \rho=\rho_w-\rho_{{\rm I}h}$,
is correctly predicted by SCAN as 0.086~g/mL,
while the opposite sign 
is predicted by PBE.
SCAN also correctly predicts $\rho_w/\rho_{{\rm I}h}\simeq1.089$,
in agreement with the experimental value of $\simeq1.087$,
and in contrast to the 0.908 ratio via PBE.
{Note that the water density obtained by PBE is slightly lower
than the results of previous studies and we discuss the difference in the SI.}

\section{Tetrahedral Structure of the H-bond Network}
With the translational order encoded in RDFs
well captured by SCAN, we now focus on the tetrahedral orientational ordering of liquid water induced by the
H-bond network.
An ideal tetrahedral H-bonding structure shown in the inset of Fig.~3(a)
is formed because a water molecule can possess four optimal H-bonds:
two accepting and two donating.
Thermal fluctuations break and reform H-bonds, causing
the tetrahedral structures in liquid water to be distorted or broken by entropic effects.
This, combined with the increased packing due to vdW interactions,
leads to an average number of H-bonds per molecule slightly less than four in liquid water.
To illustrate the impact of SCAN on the H-bond network, distributions of the number of
H-bonds per molecule are presented in Fig.~3(a).
The percentage of water molecules participating in four H-bonds drops from 72\% in PBE to 56\% in SCAN.
This suggests that H-bonds are weaker with SCAN than with PBE.
SCAN predicts an average of 3.61 H-bonds per molecule,
smaller than the 3.77 obtained from PBE.
This reduction in the number of H-bonds
is consistent with the influences of the underlying SCAN functional on liquid water:
directional H-bonds are weakened and more easily broken by thermal fluctuations.
The increased disorder is further stabilized
by the inclusion of the intermediate-ranged vdW interactions naturally arising in SCAN.
The reduction of H-bonds produced by SCAN disrupts the tetrahedral structure of liquid water.
To quantify the amount of tetrahedral order, we adopt the tetrahedral order parameter
$q$~\cite{14JCP-Rob}.
A perfect tetrahedral local environment corresponds to $q=1$,
and $q$ decreases as the local structure becomes less tetrahedral.
Following experimental work~\cite{08L-Soper}, we evaluate $q$ using a cutoff radius
that yields an average coordination number of 4.
The resulting cutoffs
are 3.15 and 3.45 \AA~for SCAN and PBE, respectively, with SCAN in better agreement with
the cutoff of 3.18~\AA~inferred from experiment~\cite{08L-Soper}.
Despite the high first peak in the PBE $g_{\mathrm{OO}}$(r),
the shorter cutoff from SCAN suggests
a more compact first coordination shell,
consistent with the higher density of liquid water it predicts.
PBE results in an overly tetrahedral liquid (Table~1).
SCAN, however, yields $q$ in better agreement with experiments on heavy water~\cite{08L-Soper},
suggesting that SCAN provides a more accurate structural description of
the fluctuating H-bond network.
Three-body correlations in water can be quantified
by the bond angle distribution $P_{\mathrm{OOO}}(\theta)$,
where $\theta$ is the angle formed by an oxygen of a water molecule and two of its oxygen neighbors;
neighbors are defined using the same cutoff as above~\cite{08L-Soper}.
The PBE $P_{\mathrm{OOO}}$ in Fig.~3(b) displays a high peak centered around the tetrahedral angle, 109.5$^{\circ}$
and is a much narrower distribution than that from experiment.
This indicates that PBE
overestimates the tetrahedral character of the liquid, consistent with the above-described overstructuring.
In stark contrast, the SCAN $P_{\mathrm{OOO}}$ is in excellent agreement with experiment,
with almost exactly the same widths and intensities of the two peaks close to 109.5$^{\circ}$ and 55$^{\circ}$, Fig.~3(c).
The peak located near 109.5$^{\circ}$ arises from tetrahedral structures.
The peak at $\theta\simeq55^{\circ}$ is related to broken H-bonds and interstitial, non-H-bonded water,
and major differences between SCAN and PBE are observed in this region of the distribution.
We decompose $P_{\mathrm{OOO}}(\theta)$ into three contributions
according to the number of H-bonds a water molecule formed within a water triplet.
The $P_{\mathrm{OOO}}(\theta)$ of triplets formed with 2, 1, and 0 H-bonds
are plotted in Figs.~3(b) and~3(c).
Triplets involving 2 H-bonds dominate the PBE $P_{\mathrm{OOO}}(\theta)$,
while triplets with broken (0 or 1) H-bonds contribute much less.
In contrast, triplets with less than 2 H-bonds contribute significantly to the
$P_{\mathrm{OOO}}(\theta)$ predicted by SCAN, especially near $\theta\approx 55^{\circ}$.
The free energy as a function of $\theta$ and the distance $r$
between neighboring oxygen atoms in the triplet
reveals additional insights, 
as shown in Figs.~3(d) and~3(e).
As expected, the minimum free energy corresponds to tetrahedral-like structures with $\theta\simeq109.5^{\circ}$
and $r\approx2.7$~\AA.
In contrast to PBE, SCAN predicts a significant fraction of triplets with
$\theta$ far from $109.5^{\circ}$, indicating that the SCAN liquid is more disordered.
The free energies suggest that
water molecules in the first coordination shell
experience a smaller free energy barrier to
adopt a broad range of $\theta$-values with SCAN than with PBE.
Importantly, there are substantial differences between the two functionals
in describing the dependence of the free energy on $r$.
With PBE, as $r$ is increased away from the free energy minimum,
$\theta$ hardly moves from 109.5$^{\circ}$, as depicted by the red arrow in Fig.~3(d).
This is consistent with the over-structuring of water by PBE
and implies that $\theta$ is weakly influenced by fluctuations of the first coordination shell.
In contrast, SCAN produces a stronger correlation between $r$ and $\theta$,
such that the free energy is lowered at larger $r$ by decreasing $\theta$,
illustrated by the red arrow in Fig.~3(e).
This is consistent with the higher population of non-H-bonded, interstitial waters in the SCAN prediction.
These non-tetrahedrally oriented water molecules contribute significantly to
$P_{\mathrm{OOO}}(\theta)$ below $109.5^{\circ}$ and highlight the reduced tetrahedrality
of the SCAN H-bond network.

\section{Dynamics}
Changes in the H-bond energy
alter the delicate enthalpy-entropy balance in liquid water that dictates its dynamic properties;
for example,
breakage and formation of H-bonds through thermal fluctuations controls diffusion.
Thus, stronger H-bonds tilt the enthalpy-entropy balance toward energetic contributions,
reducing the tendency to break H-bonds and consequently lowering the diffusion coefficient $D$.
We estimate $D$ from the long-time limit of the mean squared displacement,
averaged over the oxygen and hydrogen atoms (see SI).
Indeed, the $D$ value of PBE is an order of magnitude smaller than that of experiment,
while SCAN improves the estimate of $D$ to near agreement with experiment.
%
%

H-bond dynamics are more directly probed via the second-order
rotational correlation function of the O-H bond vector $\mathbf{r}_{\mathrm{OH}}$,
$C_2(t)=\langle P_2(\mathbf{r}_{\mathrm{OH}}(t)\cdot\mathbf{r}_{\mathrm{OH}}(0))\rangle/\langle P_2(\mathbf{r}_{\mathrm{OH}}(0)^2)\rangle$,
where $P_2(x)$ is a second-order Legendre polynomial.
The integral of $C_2(t)$ yields the rotational correlation time $\tau_2$ of the O-H bond;
correlation functions and details surrounding $\tau_2$ computation are given in the SI.
SCAN predicts a value of $\tau_2$ in agreement with
nuclear magnetic resonance spectroscopy~\cite{01JACS-Ropp}, Table~1,
while rotational dynamics are slowed in the PBE system.
The mechanism for rotational relaxation of the O-H bond vector is associated with breaking a H-bond.
In PBE, H-bonds are too strong, significantly hindering this pathway.
SCAN appropriately predicts the weight of these pathways for rotational relaxation due to its accurate description of H-bonding.

\section{Conclusions and Outlook}
The SCAN density functional provides
a genuinely predictive {\it ab initio} model of liquid water.
Importantly, SCAN is a long-awaited exchange-correlation functional
that can correctly predict liquid water that is denser than ice at ambient conditions.
SCAN excellently describes covalent and H-bonds due to an improved description of electronic structure,
and captures intermediate-ranged vdW interactions
that further improve the structure and thermodynamics of liquid water.
These vdW forces can play a critical and active role at interfaces,
for example, underlying drying transitions~\cite{Baer:2011rz,11JSP-Remsing,13JPCB-Remsing},
instilling confidence that SCAN will enable predictive modeling of heterogeneous chemical environments.
However, there are still improvements to be made regarding the water structure.
SCAN predicts a slightly over-structured first peak of g$_{\mathrm{OO}}$(r).
Previous studies have attributed the over-structuring
to self-interaction errors~\cite{81B-Perdew}, which can be mitigated by including
a fraction of exact exchange in hybrid functionals.
Moreover, the first peak in g$_{\mathrm{OH}}$(r) is too narrow,
and the error is dominated by the lack of NQEs of hydrogen~\cite{08L-Morrone}.
The widths and intensities of peaks in the computed DOS
are also respectively narrower and higher than those in the experimental DOS.
In fact, DFT is not rigorous for photoemission spectra, and
does not include lifetime broadening;
NQEs, however, can additionally broaden the DOS
bringing the resulting widths and intensities in closer agreement with experiment~\cite{16L-WeiChen}.
NQEs can be accounted for within the Feynman discretized path-integral approach~\cite{08L-Morrone,13PNAS-Ceriotti,16L-WeiChen}.

In conclusion, the SCAN XC functional within DFT shows promising predictive power
and will likely enable confident {\it ab initio} predictions for complex systems
at the forefront of physics, chemistry, biology, and materials science~\cite{13PNAS-Klein}.

\section{Methods}
We performed Car-Parrinello molecular dynamics~\cite{85L-CPMD} in \textsc{Quantum ESPRESSO}~\cite{09JPCM-QE}.
We employed the Hamann-Schl\"{u}ter-Chiang-Vanderbilt pseudopotentials~\cite{79L-Hamann}
generated using PBE. The valence electrons, including
the $1s$ electron of H and the $2s^2p^4$ electrons of O, were treated explicitly.
The energy cutoff was 150 Ry.
Simulations were performed in the isothermal-isobaric ensemble (constant {\it NpT}) by using the Parrinello-Rahman barostat~\cite{80L-Parrinello}
and a single Nos\'{e}-Hoover thermostat~\cite{92JCP-Martyna} with a frequency of 60~THz
to maintain a constant pressure ($p$) and temperature ($T$), respectively.
$T=330$~K for liquid water and 273~K for ice I{\it h};
the 30~K increase above ambient conditions in the former mimics NQEs on the liquid structure~\cite{08L-Morrone}.
We adopted a cubic cell with $N$=64 water molecules.
The fictitious mass of the electrons was set to 100~au and the corresponding
mass pre-conditioning with a kinetic energy cutoff of 25 Ry was used
to all Fourier components of wavefunctions~\cite{94B-Tassone}.
The deuterium mass was used instead of hydrogen to enable the use of a timestep of 2~au;
dynamics were compared to D$_{2}$O instead of H$_{2}$O.
SCAN and PBE trajectories for water were 30.0~and 20.0~ps in length, respectively.
Corresponding trajectories for ice I{\it h} were {11.1 and 13.8~ps}, respectively.
The first 5~ps of each trajectory was used for equilibration and the remainder used for analysis.
We utilized a standard geometric criterion for hydrogen bonding;
covalently bonded O-H are associated with an O-H distance less than 1.24~\AA~and
H-bonds have an O-O distance less than 3.5~\AA~and a $\angle{\rm OOH}$ angle less than $30^{\circ}$~\cite{96N-Chandler}.
Additional details are in the SI.

\newpage

\section{Author contributions}
X.W. and M.C. designed research. M.C. and Z.S. performed {research}.
H.-Y.K., M.F.C.A., and B.S. {contributed new methods}.
M.C. and R.C.R. analyzed data.
All authors contributed to {writing the paper}.

\section{Acknowledgements}
This work was supported by the U.S. Department of Energy (DOE) SciDac under Grant
\#DE-SC0008726. This research used resources of the National
Energy Research Scientific Computing Center, a DOE Office of Science User Facility
supported by the Office of Science of the U.S. DOE under
Contract \#DE-AC02-05CH11231. RCR, ZS, and JPP were supported as
part of the Center for the Computational Design of Functional Layered Materials, an Energy Frontier Research Center funded by the U.S. DOE, Office of Science, Basic Energy Sciences under Award \#DE-SC0012575.
MFCA is partially supported by the CNPq - Brazil.
XW is partially supported by the National Science Foundation (NSF), DMR
under Award \#DMR-1552287.

\newpage

\section{Supporting Information}
\subsection{{Computational Details}}
The {\it NpT} algorithm was implemented in the {\sc Quantum ESPRESSO}~\cite{09JPCM-QE} package.
In our water simulations, all of the plane waves $\{\mathbf{G}\}$ with kinetic energies below 150 Ry were included, {and we followed Ref.~\cite{bernasconi}}
to maintain a constant plane wave
kinetic energy cutoff of $E_{0}$=130 Ry for a fluctuating cell
{by adding a smooth step function with
height $A=200$ Ry and width $\sigma=15$ Ry to the plane wave kinetic
factor as $\mathbf{G}^{2} \rightarrow \mathbf{G}^{2} + A[1+\mathrm{erf}(\frac{\mathbf{G}^2/2-E_0}{\sigma})]$,
where $erf$ is the error function.
The reference cells were chosen to be cubes with side lengths of 14.3345 \AA~and
12.6579 \AA~for PBE- and SCAN-based AIMD simulations, respectively.}

{We ran parallel AIMD simulations using both
strongly constrained and appropriately normed (SCAN)~\cite{15L-Sun}
and Perdew-Burke-Ernzerhof (PBE)~\cite{96L-PBE} exchange-correlation functionals
on 216 computer cores and recorded the wall times of 100 MD steps.
The system was bulk liquid water consisting of 64 water molecules as utilized in this work.
Both simulations were carried out on nodes with 2 x Intel Ivy Bridge @ 2.4 GHz and up to 64 GB RAM.
We obtained 6.44 and 3.89 seconds/step for SCAN and PBE functionals, respectively,
with SCAN being only 1.66 times more costly than PBE.
Therefore, we conclude the cost of using the SCAN functional in studying liquid water is
not dramatically more expensive than that of using the PBE functional
and can be considered on the same level.}

\subsection{Bulk Densities}
Bulk densities as a function of time are shown in Fig. 4 for the
{SCAN and PBE} descriptions
of liquid water and ice I{\it h}.
The ice I{\it h} phase remained solid throughout and did not transform to the liquid phase in the AIMD trajectories.
{The averaged densities are listed in Table 1 in the main text with
the errors bars corresponding to one standard deviation.}
With PBE the dynamics of ambient liquid water is very sluggish
and we find that at 330 K the mean density of liquid water
(0.85 g/mL as shown in Table 1 in the main text) computed with PBE
can vary within ~0.01 g/mL (as estimated by another independent run of
more than 60 ps by members of Roberto Car’s group with PBE and all the same parameters)
depending on the initial configuration and trajectory length.
Since PBE liquid density is well below PBE ice I{\it h} density
we did not try to reduce the statistical uncertainty in the PBE liquid density.
In addition, a recent study found structural, dynamical, and electronic properties of liquid water as
obtained by AIMD simulations utilizing {\sc CP2K}~\cite{CP2K} and  {\sc Quantum ESPRESSO} packages~\cite{16JCTC-Miceli}compared well,
with the latter equilibrium density 0.02 g/mL lower, and $g_{\rm OO}(r)$ first-peak ~0.1 higher, than corresponding CP2K values.
These differences are comparable to the statistical uncertainties.
Therefore the small difference in the PBE liquid density found here: 0.85 versus 0.865-0.887 g/mL by~\cite{09JPCB-Schmidt},
as well as the slightly higher $g_{\rm OO}(r)$ first-peak: 3.61 versus 3.36-3.54~\cite{09JPCB-Schmidt}
is attributed to both statistical uncertainties and numerical differences of these approaches.
The densities clearly fluctuate around an average value for each trajectory, illustrating equilibration of the trajectories.
Moreover, the fact that water is denser than ice I{\it h} in the SCAN prediction is clearly observed,
while the opposite is found for PBE.
Finally, we note that fluctuations in the density of water are larger in the SCAN trajectory than in the PBE
trajectory.
This indicates that water is more compressible in the SCAN description than PBE, which produces
a more rigid and ordered liquid structure.

\begin{figure}[tbh]
\label{fig3}
\begin{center}
\includegraphics[width=0.43\textwidth]{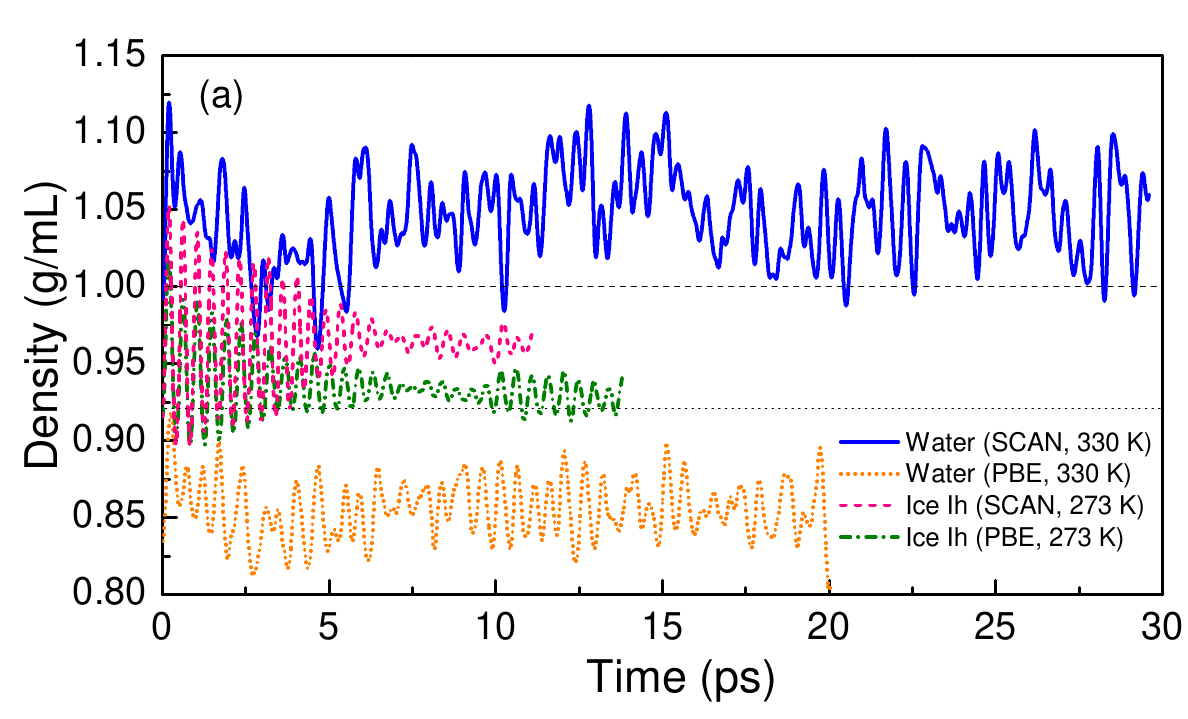}
\end{center}
\caption
{Density fluctuations of liquid water and ice I{\it h}
as obtained from both SCAN and PBE trajectories using the isobaric-isothermal ensemble ({\it NpT}).
A relatively shorter trajectory was generated for ice I{\it h}
because its density of solid phase converges quickly.
The PBE functional incorrectly predicts that ice I{\it h} (green line) is denser than water (orange line),
while the SCAN functional successfully captures the larger density of water (blue line) than that of ice I{\it h} (pink line).
The black dashed and dotted lines represent the approximate experimental values of water density
at ambient conditions (300 K)
and ice density at 273 K under ambient pressure. An additional 30 K was applied to water to
mimic the nuclear quantum effect~\cite{08L-Morrone}.
The averaged densities are listed Table 1 in the main text.
}
\end{figure}

\subsection{van der Waals Interactions in SCAN and PBE}
As discussed extensively in the main text,
an accurate description of water and ice from first principles is challenging
because the H-bond network of water arises from a delicate balance of strong intra-molecular covalent bonds,
weak inter-molecular H-bonds, and even weaker vdW interactions.
GGA functionals exhibit delocalization problems and
the intermediate- and long-ranged vdW attraction is strongly underestimated.
To be specific, the exchange energy density obtained from GGA
is much more negative than LDA in regions with a large reduced density gradient.
Therefore, the attractive vdW interactions between water molecules are missing in GGAs,
which results in more ordered water molecules and a lower bulk density.
The SCAN functional captures this delicate balance between covalent bonds, H-bonds, and vdW interactions in water,
which is critical in accurately describing the water density.

By including a dimensionless variable $\alpha$ in the kinetic energy density,
SCAN can reduce to different GGAs by recognizing
covalent single bonds when $\alpha=0$,
slowly varying densities when $\alpha\simeq1$,
and non-covalent bonds when $\alpha>1$.
It was recently demonstrated that SCAN captures the intermediate-ranged vdW interactions
for a variety of materials~\cite{16NC-Sun}.
To further illustrate this point,
we applied the Tkatchenko-Scheffler (TS) vdW scheme~\cite{09vdW-TS} to both the PBE and SCAN functionals.
The TS scheme determines the "turn-on" radius for atom pairs based on the XC functional used,
and a larger TS scaling parameter $S_R$ implies that a larger radius is adopted to turn on the vdW interactions.
The scaling parameters $S_R$ were obtained by fitting to the S22 molecular database.
We find that in both water and ice structures, the scaling parameters are 0.94 and 1.17 for the PBE and SCAN functionals, respectively.
The 24.5\% larger scaling parameter in SCAN demonstrates that it captures the vdW interactions out to significantly larger distances than PBE.

\subsection{Mean Squared Displacements}

\begin{figure}[tb]
\label{figs2}
\begin{center}
\includegraphics[width=0.35\textwidth]{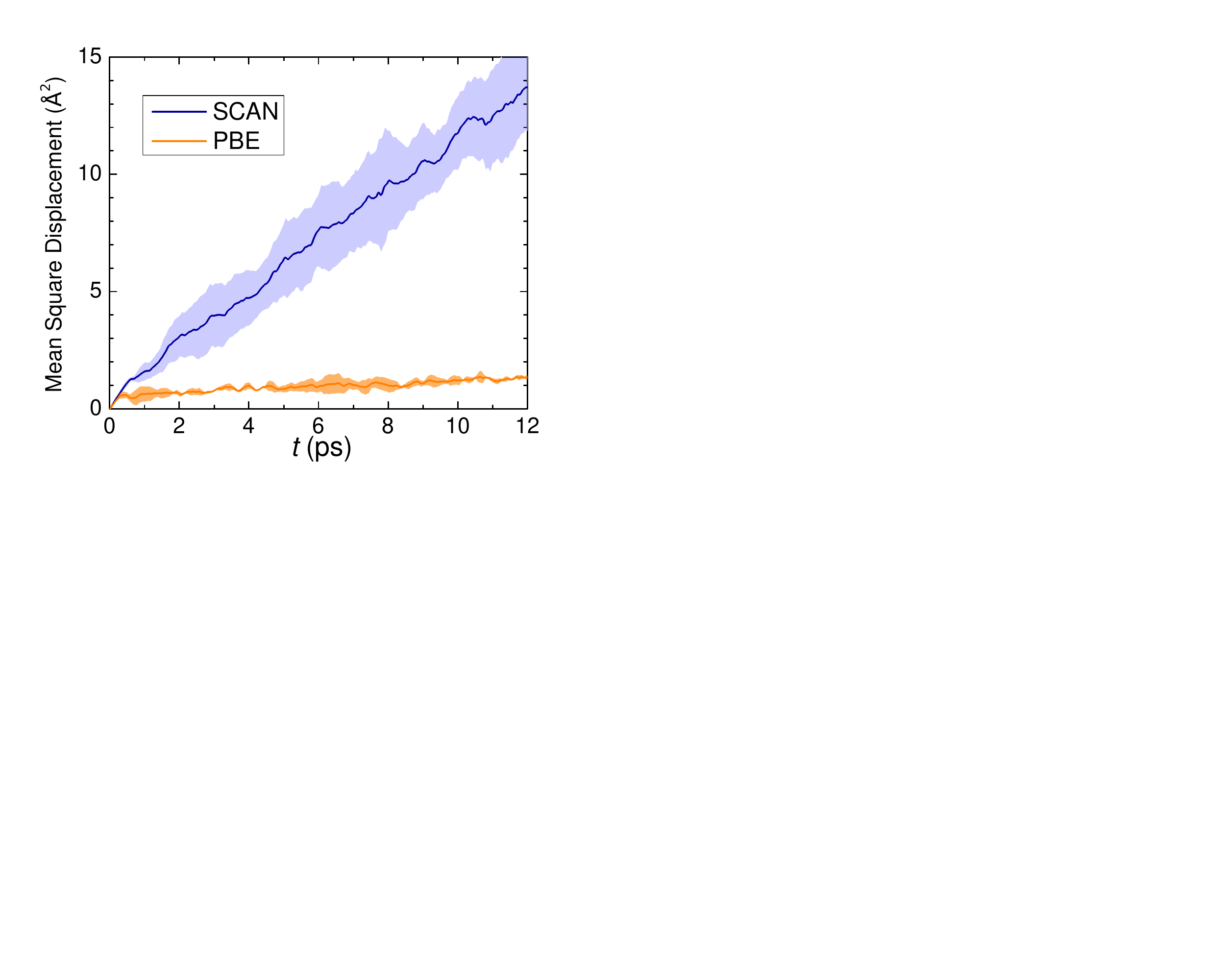}
\end{center}
\caption
{Mean squared displacements for the SCAN and PBE systems consisting of 64 water molecules.
Shaded regions indicate one standard error.
}
\end{figure}

The mean squared displacements (MSDs) computed from SCAN and PBE trajectories
are shown in Fig. 5, where the center-of-mass positions of water molecules
were used to compute MSDs.
Each trajectory was divided into sections to compute the MSDs, each 12~ps in length and separated by 3.0~ps.
We chose five and three sections for SCAN and PBE trajectories, respectively.
Next, a linear fitting of the long-time, linear region of the MSDs was performed to obtain the $D$ values. Finally, the obtained $D$ values were averaged and the result is listed in Table~1 of the main text; the slope of the linear region is equal to $6D$.
Clearly, water diffuses much faster in the SCAN description than PBE,
due to the weaker H-bonds predicted by the SCAN functional.
The computed $D$ from SCAN and PBE are 0.190 and 0.018 $\AA^2/ps$, respectively.
The diffusion coefficient from PBE is close to the
0.020 $\AA^2/ps$ reported in a previous work~\cite{14JCP-Rob}.
However, we note that correlations may exist in divided sections and
affect the accuracy of computed diffusion coefficients~\cite{04JPCB-Kuo,06JCTC-Kuo}.
More accurate diffusion coefficients require longer simulation times
and we leave this investigation for future work.

\subsection{Rotational Time Correlation Functions}

\begin{figure}[tbh]
\label{figs2}
\begin{center}
\includegraphics[width=0.49\textwidth]{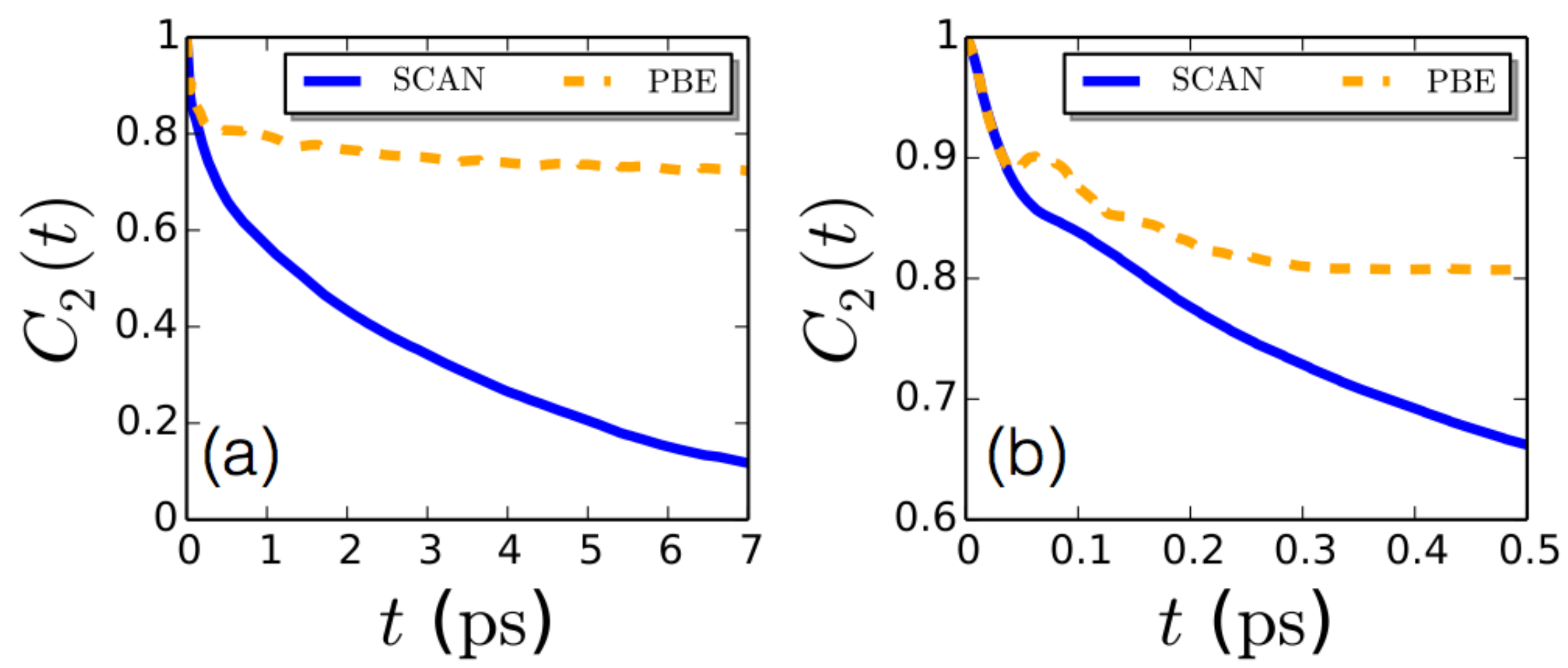}
\end{center}
\caption
{Second-order rotational time correlation functions, $C_2(t)$,
for the O-H bond vector of water as described by SCAN and PBE.
Panel (b) is the same as (a), but zoomed in on the region from 0 to 0.5 ps.
}
\end{figure}

\begin{figure}[tb]
\label{fig3}
\begin{center}
\includegraphics[width=0.43\textwidth]{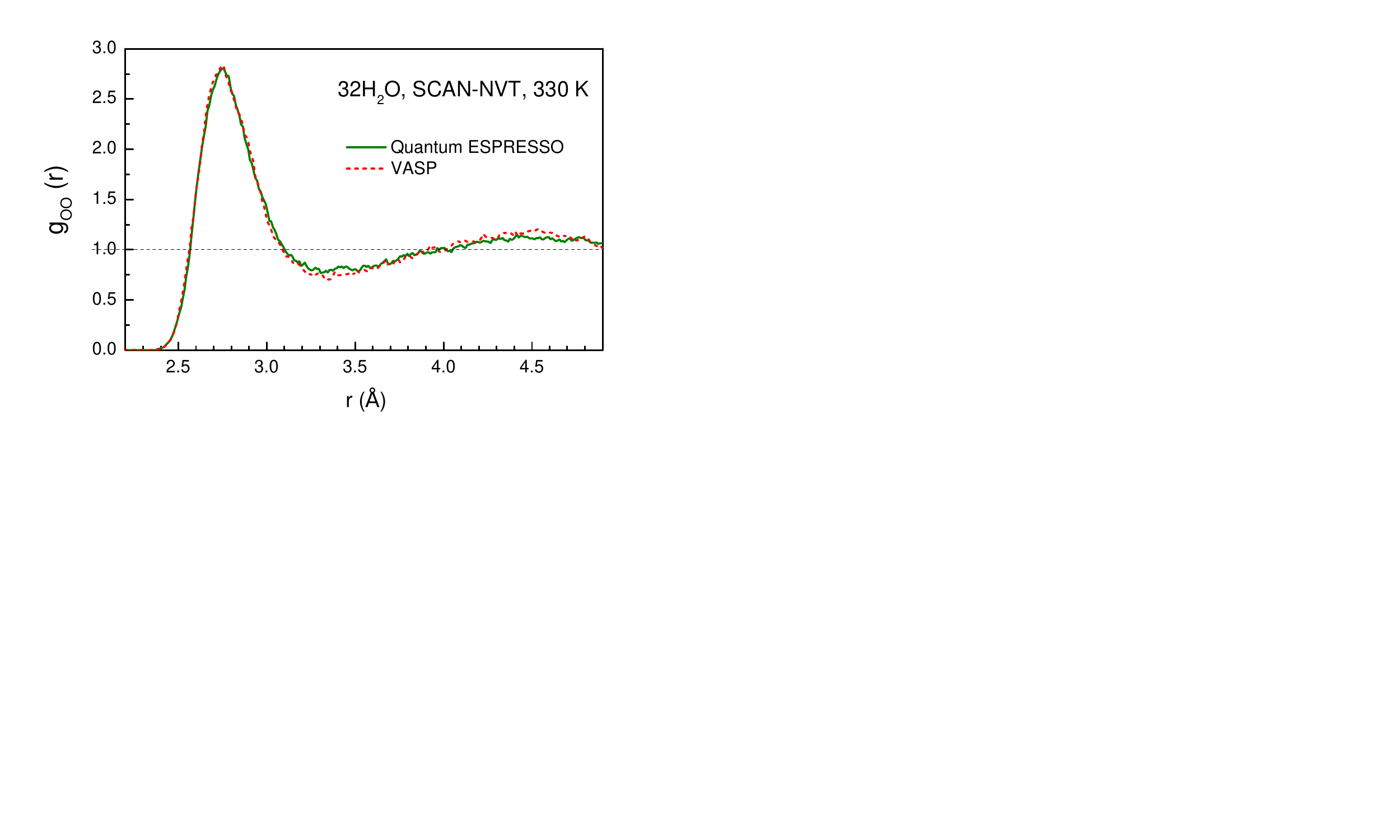}
\end{center}
\caption
{Oxygen-oxygen radial distribution functions
$g_{\mathrm{OO}}(r)$ for 32 water molecules in condensed phase
as obtained from
the SCAN functional implemented in VASP and {\sc Quantum ESPRESSO}
electronic structure packages.
}
\end{figure}

The second-order rotational correlation function for the O-H bond vector $\mathbf{r}_{\mathrm{OH}}$
was calculated according to
$C_2(t)=\langle P_2(\mathbf{r}_{\mathrm{OH}}(t)\cdot \mathbf{r}_{\mathrm{OH}}(0))\rangle/\langle P_{2}(\mathbf{r}_{\mathrm{OH}}(0)^2)\rangle$,
where $P_2(x)$ is a second-order Legendre polynomial.
These time correlation functions as predicted for water by PBE and SCAN are shown in Fig. 6.
Clearly, SCAN results in significantly faster rotational dynamics,
evidenced by the much faster decay of the SCAN $C_2(t)$ than that of PBE.
As discussed in the main text, this is because SCAN results in weaker and more physical hydrogen bonding interactions than PBE.

The short-time behavior of $C_2(t)$ is shown in Fig. 6(b).
We find that the initial short-time decay ($<$50 fs) of $C_2(t)$ is identical in the two models.
This initial decay is due to rapid inertial,
librational motions of water that do not require hydrogen bond breakage.
Thus, in both models, hydrogen bonds are still intact on this short timescale and no major differences between SCAN and PBE are found.
However, the oscillation in $C_2(t)$ occurs at different times in the two models.
A larger oscillation is found in PBE, which occurs prior to the slight oscillation in the SCAN $C_2(t)$.

Finally, to estimate the rotational correlation time $\tau_2$ of the O-H bond,
we must integrate $C_2(t)$. To do so, we first note that the long time decay of $C_2(t)$
is well described by an exponential.
Thus, we fit $C_2(t)$ to an exponential at long times (after the initial change in slope associated with librations).
The fitted exponential is then used to describe the decay of $C_2(t)$ for times longer than 7 and 9 ps in PBE and SCAN, respectively.
We then numerically integrate this composite $C_2(t)$ to obtain $\tau_2$, which are listed in Table 1 of the main text.

\subsection{Validation of SCAN in Different Packages}
To further validate our results employing the SCAN functional for liquid water,
we ran AIMD simulations on
a cell of 32 water molecules for 20 ps by employing both Vienna Ab initio Simulation Package
(VASP)~\cite{96B-Kresse} and
{\sc Quantum ESPRESSO}~\cite{09JPCM-QE} packages.
The cell was chosen to be a cube of side length 9.877 \AA.
We adopted the NVT ensemble with the Nos\'{e}-Hoover thermostat and
the temperature was set to 330 K. We used the mass of deuterium instead of hydrogen to speed up the convergence.
In VASP, we used projector-augmented-wave (PAW) potentials
with configurations of [O]2s$^2$2p$^4$ and [H]1s$^1$. In particular,
we chose the hard PAW potentials for oxygen and hydrogen atoms
and set the energy cutoff to 1200 eV in order to converge our results.
The Born-Oppenheimer molecular dynamics
was performed with a time step of 0.5 fs.
In {\sc Quantum ESPRESSO}, we carried out Car-Parrinello molecular dynamics~\cite{85L-CPMD}
and the settings were chosen to be the same as described in the main text.

In Fig. 7, the oxygen-oxygen radial distribution functions
$g_{\mathrm{OO}}(r)$ are shown for the above two calculations.
We find both electronic structure packages yield almost the same $g_{\mathrm{OO}}(r)$ features.
In particular, the first peak from both packages are almost identical.
The results suggest that the properties of liquid water
as predicted by the SCAN functional are reliable and are reproducible
with converged basis set and electron dynamics.

\subsection{Radial Distribution Function $g_{OH}$}
We plot in Fig. 8 the zoomed-in first peak of radial distribution
function $g_{OH}$ from both PBE- and SCAN-based AIMD simulations.
The first peak position represents the length of O-H covalent bond and
we can see that SCAN predicts a slightly shorter covalent bond (0.977 \AA)
than that from PBE (0.989 \AA).

\begin{figure}[tbh]
\label{fig5}
\begin{center}
\includegraphics[width=0.35\textwidth]{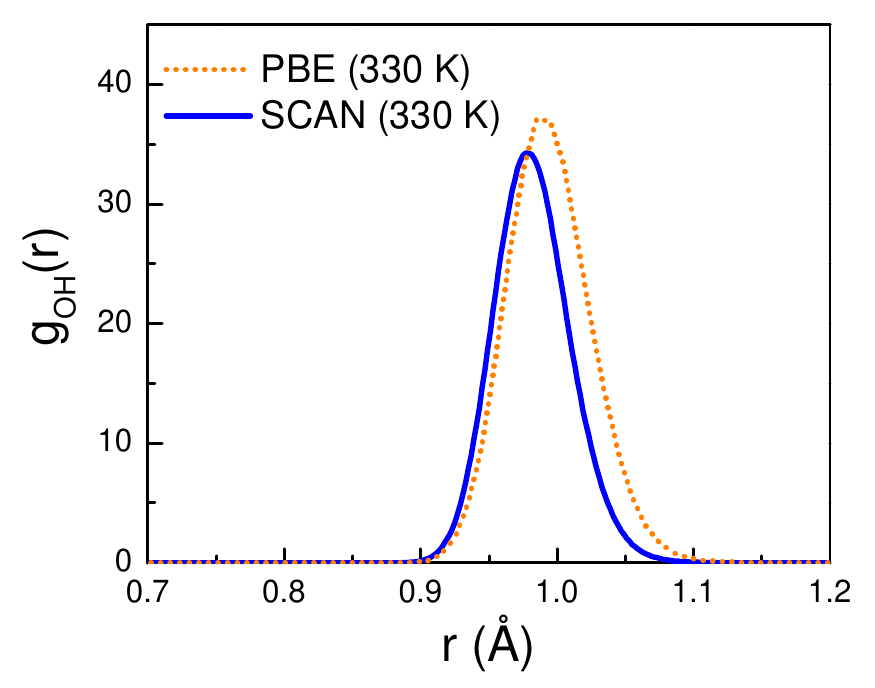}
\end{center}
\caption
{Zoomed-in radial distribution function $g_{OH}$ (as shown in Fig. 1(b) in the main text)
as obtained from SCAN- and PBE-based AIMD simulations.
}
\end{figure}

\newpage

\bibliography{references}

\end{document}